\documentclass{elsarticle}

\usepackage{algorithm,multirow,rotating}
\usepackage{algorithmic}

\begin{document}

\begin{frontmatter}

\title{A note on a sports league scheduling problem}

\author{Jean-Philippe Hamiez\corref{cor}}
\ead{hamiez@info.univ-angers.fr}
\cortext[cor]{Principal corresponding author. Phone: (+ 33) 241 735 385. Fax: (+ 33) 241 735 073.}

\author{Jin-Kao Hao}
\ead{hao@info.univ-angers.fr}
\ead[url]{www.info.univ-angers.fr/pub/hao}

\address{Universit\'e d'Angers -- LERIA, 2 boulevard Lavoisier, 49045 Angers CEDEX 01, France}

\begin{abstract}
Sports league scheduling is a difficult task in the general case. In this short note, we report two improvements to an existing enumerative search algorithm for a NP-hard sports league scheduling problem known as ``prob026'' in CSPLib. These improvements are based on additional rules to constraint and accelerate the enumeration process. The proposed approach is able to find a solution (schedule) for all prob026 instances for a number $T$ of teams ranging from 12 to 70, including several $T$ values for which a solution is reported for the first time.
\end{abstract}


\end{frontmatter}

\section{Introduction}
\label{SectIntroduction}

The sports league scheduling problem studied in this note, called ``prob026'' in CSPLib \cite{CSPLib} and also known as the ``balanced tournament design'' problem in combinatorial design theory \cite[pages 238-241]{Colbourn&Dinitz1996}, is a NP-hard problem \cite{Briskorn&al2010} that seems to be first introduced in \cite{Gelling&Odeh1974}:

\begin{itemize}
\item There are $T=2n$ teams (i.e., $T$ even). The season lasts $W=T-1$ weeks. Weeks are partitioned into $P=T/2$ slots called ``periods'' or ``stadiums''. Each week, one match is scheduled in every period;
\item $c_\mathcal{H}$ constraint: All teams play each other exactly once ($\mathcal{H}$alf competition);
\item $c_\mathcal{P}$ constraint: No team plays more than twice in a $\mathcal{P}$eriod. This constraint may be motivated by the equal distribution of stadiums to teams;
\item $c_\mathcal{W}$ constraint: Every team plays exactly one game in every $\mathcal{W}$eek of the season, i.e., all teams are different in a week.
\end{itemize}

The problem then is to schedule a tournament with respect to these definitions and constraints. A solution to prob026 is a complete assignment of $D = \{(t,t'), 1 \le t < t' \le T\}$ items (couples of $t$eams) to variables of $X=\{x=\langle p,w \rangle, 1 \le p \le P, 1 \le w \le W\}$ (couples of $p$eriods and $w$eeks) verifying the constraint set $C=\{c_\mathcal{H}, c_\mathcal{P}, c_\mathcal{W}\}$, $\langle p,w \rangle = (t,t')$ meaning that team $t$ meets team $t'$ in period $p$ and week $w$. Thus, a solution can be conveniently represented by a $P \times W$ sized table, whose items are integer couples $(t, t')$, see Table~\ref{Example_valid_schedule} for an example of a valid schedule for $T = 8$. For $T = 70$ teams, this represents a problem with 2\,415 variables and 2\,415 values per variable. There are $T(T - 1)/2$ matches to be scheduled. A valid schedule can be thought of as a particular permutation of these matches. So, for $T$ teams, the search space size is $[T(T - 1)/2]!$.

\begin{table}[h]
\begin{center}
\caption{A valid schedule for 8 teams.}
\label{Example_valid_schedule}
\begin{tabular}{cccccccc}
\noalign{\smallskip}\hline
\multirow{2}{*}{Periods} & \multicolumn{7}{c}{Weeks} \\
\cline{2-8}
	& 1 	& 2 	& 3 	& 4 	& 5		& 6		& 7\\
\hline
1	& 1,2 & 6,8 & 2,5 & 4,5 & 4,7 & 3,8 & 1,7\\
2	& 3,7 & 5,7 & 3,4 & 1,8 & 5,6 & 2,4 & 2,6\\
3	& 4,6 & 1,4 & 7,8 & 3,6 & 2,8 & 1,5 & 3,5\\
4	& 5,8 & 2,3 & 1,6 & 2,7 & 1,3 & 6,7 & 4,8\\
\hline
\end{tabular}
\end{center}
\end{table}

Direct construction methods exist when $(T-1) \bmod 3 \neq 0$ \cite{Hamiez&Hao2004a,Haselgrove&Leech1977} or $T/2$ is odd \cite{Lamken&Vanstone1985,Schellenberg&al1977}. However, finding a solution (schedule) in the general case for any arbitrary $T$ remains a highly challenging task. Indeed, to our knowledge, the best performing search algorithm \cite{Hamiez&Hao2008} can solve all the instances for $T$ up to 50, but only some cases when $50 < T \le 70$. Other representative solution approaches include integer programming \cite{McAloon&al1997} (limited to $T \le 12$), transformation into the SAT problem \cite{Bejar&Manya2000} ($T \le 20$), distributed approach ($T \le 28$ according to \cite{Gomes&al1998a}), constraint programming \cite{vanHentenryck&al1999} and tabu search \cite{Hamiez&Hao2001} ($T \le 40$).

In this paper, we present two improvements to the \texttt{En}umera\-tive \texttt{A}lgorithm (\texttt{EnASS}) proposed in \cite{Hamiez&Hao2008}. With the proposed enhancements, \textbf{all} the instances for $12 \leq T \le 70$ can now be solved.

We provide in the next section a brief recall of the original \texttt{EnASS} method. We show then in the following sections a new \texttt{EnASS} variant that solves \textbf{all} instances up to $T = 60$ (including the problematic $T \bmod 4 = 0$ cases) and another new  variant that solves all the $12 \leq T \leq 70$ instances.

\section{A brief recall of the \texttt{EnASS} algorithm}

\texttt{EnASS} starts with a complete $\overline{s}$ conflicting assignment. $\overline{s}$ is built, in linear-time complexity, to satisfy the $c_\mathcal{W}$ and $c_\mathcal{H}$ constraints (thanks to patterned one-factorization \cite[page 662]{Colbourn&Dinitz1996}). At this stage, the remaining $c_\mathcal{P}$ constraint is not verified in $\overline{s}$, see Table~\ref{Initial_schedule_8} where team 8 appears more than twice in the 4th period.

\begin{table}
\begin{center}
\caption{Initial conflicting $\overline{s}$ schedule for 8 teams.}\label{Initial_schedule_8}
\begin{tabular}{cccccccc}
\noalign{\smallskip}\hline
\multirow{2}{*}{Periods} & \multicolumn{7}{c}{Weeks} \\
\cline{2-8}
 & 1 & 2 & 3 & 4 & 5 & 6 & 7 \\
\hline
1 & 1,2 & 2,3 & 3,4 & 4,5 & 5,6 & 6,7 & 1,7 \\
2 & 3,7 & 1,4 & 2,5 & 3,6 & 4,7 & 1,5 & 2,6 \\
3 & 4,6 & 5,7 & 1,6 & 2,7 & 1,3 & 2,4 & 3,5 \\
{\bfseries 4} & 5,{\bfseries 8} & 6,{\bfseries 8} & 7,{\bfseries 8} & 1,{\bfseries 8} & 2,{\bfseries 8} & 3,{\bfseries 8} & 4,{\bfseries 8} \\
\hline
\end{tabular}
\end{center}
\end{table}

\begin{algorithm}
\caption{\texttt{EnASS}: An overview.}
\label{AlgoEnASS}
\begin{algorithmic}[1]
\REQUIRE Two periods ($p$ and $\overline{p}$) and a week ($w$)
\IF[A solution is obtained since all periods are filled and valid according to $\mathcal{R}$]{$p=P+1$}
	\RETURN \TRUE
\ENDIF
\IF[Period $p$ is filled and valid according to $\mathcal{R}$, try to fill next period]{$w=w_l+1$} \label{BeginPeriodFilled}
	\RETURN \texttt{EnASS}($p+1, w_f, 1$)
\ENDIF \label{EndPeriodFilled}
\IF[Backtrack since no match from week $w$ in $\overline{s}$ can be scheduled in period $p$ of week $w$ without violating $\mathcal{R}$]{$\overline{p}=P+1$}
	\RETURN \FALSE
\ENDIF
\IF[The $\overline{s}\langle \overline{p},w \rangle$ match is already scheduled, try next match]{$\exists\,1 \le p' < p : \langle p',w \rangle =\overline{s}\langle \overline{p},w \rangle$}
	\RETURN \texttt{EnASS}($p, w, \overline{p}+1$)
\ENDIF
\STATE $\langle p,w \rangle \gets \overline{s}\langle \overline{p},w \rangle$ \label{Assign}
	\COMMENT{Schedule the $\overline{s}\langle \overline{p},w \rangle$ match in period $p$ of week $w$}
\IF[The previous assignment and next calls lead to a solution]{$\mathcal{R}$ is locally verified and \texttt{EnASS}$(p, w+1, 1)=$ \TRUE}
	\RETURN \TRUE
\ENDIF
\STATE \COMMENT{From this point, $\mathcal{R}$ is locally violated or next calls lead to a failure}
\STATE Undo step \ref{Assign} \COMMENT{Backtrack}
\RETURN \texttt{EnASS}($p, w, \overline{p}+1$) \COMMENT{Try next value for $\langle p,w \rangle$}
\end{algorithmic}
\end{algorithm}

Roughly speaking, \texttt{EnASS} uses $\overline{s}$ to search for a valid tournament by filling a $P \times W$ table (initially empty) row by row, see Algorithm~\ref{AlgoEnASS} where $w_f$ and $w_l$ are the $f$irst and $l$ast weeks \texttt{EnASS} considers when filling any period $p$ ($1 \le w_f < w_l \le W$), $\overline{s}\langle \overline{p},w \rangle$ is the match in $\overline{s}$ scheduled in period $\overline{p}$ and week $w$, and $\mathcal{R}$ is a set of properties (or ``$\mathcal{R}$equirements'') that (partial or full) solutions must verify. \texttt{EnASS} admits three integer parameters: $p$ and $w$ specify which $\langle p,w \rangle$ variable is currently considered, $\overline{p}$ specifies the value assignment tried (see step \ref{Assign}). The function returns TRUE if a solution has been found or FALSE otherwise. Backtracks are sometimes performed in the latter case. \texttt{EnASS} is called first, after the $\overline{s}$ initialization, with $p = 1, w = w_f$ and $\overline{p} = 1$ meaning that it tries to fill the slot in the first period of week $w_f$ with the $\overline{s}\langle 1,w_f \rangle$ match.

The basic \texttt{EnASS} skeleton presented in Algorithm~\ref{AlgoEnASS} solves prob026 only up to $T = 12$ when the $\mathcal{R}$ set is restricted to $\left\{ c_\mathcal{P} \right\}$ while considering the first week as invariant with respect to $\overline{s}$ (i.e., $\forall 1 \le p \le P, \langle p, 1 \rangle = \overline{s}\langle p, 1 \rangle$) with $w_f = 2$ (since the first week is invariant) and $w_l = W$. Note that making the first week invariant helps to avoid some evident symmetries mentioned in \cite[see Sect.~4 and 5.3]{Hamiez&Hao2008}.

To tackle larger-size problems, several \texttt{EnASS} variants were considered in \cite{Hamiez&Hao2008}. \texttt{EnASS}$_0$ solved prob026 up to $T = 32$, except the $T = 24$ case, including in $\mathcal{R}$ an implicit property (called ``$c_\mathcal{D}$'' in \cite{Hamiez&Hao2008}) of all prob026 solutions: $\mathcal{R}_0 = \left\{c_\mathcal{P}, c_\mathcal{D} \right\}$. The $c_\mathcal{D}$ property was not originally mentioned in the seminal definition of the problem \cite{Gelling&Odeh1974} and seems to be first introduced in \cite{Schellenberg&al1977}. \texttt{EnASS}$_1$, derived from \texttt{EnASS}$_0$ by further including an ``implied'' requirement ($r_{\Rightarrow}$), solved all instances up to $T = 50$: $\mathcal{R}_1 = \left\{c_\mathcal{P}, c_\mathcal{D}, r_{\Rightarrow} \right\}$. Finally, \texttt{EnASS}$_2$ solved some cases (when $T \bmod 4 \neq 0$) for $T$ up to 70 with two additional invariants ($r_I$ and $r_V$): $\mathcal{R}_2 = \left\{c_\mathcal{P}, c_\mathcal{D}, r_{\Rightarrow}, r_I, r_V \right\}$.

\section{Solving all instances of prob026 up to $T = 60$}
\label{CTS60}

The rule $r'_{\Rightarrow}$ used to solve \textbf{all} prob026 instances up to $T = 60$ resembles the original $r_{\Rightarrow}$ requirement introduced in \cite[Sect.~7]{Hamiez&Hao2008}. Like $r_{\Rightarrow}$, $r'_{\Rightarrow}$ fixes more than one variable (two exactly, to be more precise) when exploring a new branch in the search tree. The difference between $r_{\Rightarrow}$ and the new $r'_{\Rightarrow}$ rule is the weeks that are concerned: While $r_{\Rightarrow}$ connects any week $w_f \le w \le P$ to week $T-w+1$, the $r'_{\Rightarrow}$ constraint links any week $1 \le w \le P - 1$ together with week $W-w+1$. More formally, $\forall\,1 \le w \le P-1, r'_\Rightarrow(p,w) \Leftrightarrow \langle p,w \rangle = \overline{s}\langle \overline{p},w \rangle \Rightarrow \langle p,W-w+1 \rangle = \overline{s}\langle \overline{p},W-w+1 \rangle$.

This leads to \texttt{EnASS}$_3$ which comes from the \texttt{EnASS}$_1$ algorithm from \cite{Hamiez&Hao2008} by replacing in $\mathcal{R}_1$ the $r_\Rightarrow$ requirement with the new $r'_{\Rightarrow}$ rule: $\mathcal{R}_3 = \{c_\mathcal{P}, c_\mathcal{D},r'_\Rightarrow\}$. Like for \texttt{EnASS}$_1$, step \ref{Assign} in the basic \texttt{EnASS} description (see Algorithm~\ref{AlgoEnASS}) may be adapted since one additional variable has now to be instantiated and $w_l$ has to be set to $P-1$ before running \texttt{EnASS}$_3$. Steps~\ref{BeginPeriodFilled}--\ref{EndPeriodFilled} have also to be modified since, when $w = w_l + 1$, the $P$ week is not yet filled (so, the $p$ period is not entirely filled either). Table~\ref{Example_valid_schedule} in Sect.~\ref{SectIntroduction} shows an example of a solution found by \texttt{EnASS}$_3$ for $T=8$: For instance, scheduling the $(3,4)$ match from week 3 in period 2 forces the $(5,6)$ match from week 5 ($5=7-3+1$) to be also in period~2.

In Table~\ref{CTS3vsCTS1}, we show for $6 \le T \le 50$ comparisons of our new  \texttt{EnASS}$_3$ variant (as well as another new  \texttt{EnASS}$_4$ variant discussed in the next section), against the \texttt{EnASS}$_1$ algorithm which solves all the instances for $T \le 50$ within 3 hours per $T$ value. The reported statistics include execution times (in seconds in all tables) and number of backtracks (columns labeled ``$|$BT$|$'') needed to find a first solution.

In Table~\ref{CTS3vsCTS2}, we show for $52 \leq T \leq 70$ comparisons between the new variant \texttt{EnASS}$_3$ (and \texttt{EnASS}$_4$) and the \texttt{EnASS}$_2$ algorithm from \cite{Hamiez&Hao2008} which solves \emph{some} instances with $T \leq 70$ where $T \bmod 4 \neq 0$. ``--'' marks in the ``Time'' (respectively ``$|$BT$|$'') columns indicate that the method found no solution within 3 hours (resp. that $|$BT$|$ exceeds the maximal integer value authorized by the compiler/system, i.e., 4\,294\,967\,295). All \texttt{EnASS} variants were coded in \texttt{C} and all computational results were obtained on an Intel PIV processor (2 Ghz) Linux station with 2 Gb RAM.

\begin{table}[h]
\begin{center}
\caption{Solving all prob026 instances up to $T=50$. }\label{CTS3vsCTS1}
\begin{small}
\begin{tabular}{rrrcrrcrr}
\noalign{\smallskip}\hline
\multirow{2}{*}{$T$}
		& \multicolumn{2}{c}{\texttt{EnASS}$_1$ \cite{Hamiez&Hao2008}}	&
		& \multicolumn{2}{c}{\texttt{EnASS}$_3$ (Sect.~\ref{CTS60})} 	&
		& \multicolumn{2}{c}{\texttt{EnASS}$_4$ (Sect.~\ref{CTS70})}\\
			\cline{2-3} \cline{5-6} \cline{8-9}
 		& Time 			& $|$BT$|$ 					&& Time 			& $|$BT$|$					&& Time		& $|$BT$|$\\
\hline
6 	& $<1$ 			& 6  								&& $<1$				& 1									&& --			& --\\
8		& $<1$ 			& 16 								&& $<1$				& 6									&& $<1$		& 5\\
10	& $<1$ 			& 715 							&& $<1$				& 350								&& --			& --\\
12 	& $<1$ 			& 86 								&& $<1$				& 25								&& $<1$		& 111\\
14 	& $<1$ 			& 451 							&& $<1$				& 65								&& $<1$		& 125\\
16 	& $<1$ 			& 557 							&& $<1$				& 713								&& $<1$		& 560\\
18 	& $<1$ 			& 1\,099 						&& $<1$				& 772								&& $<1$		& 465\\
20 	& $<1$ 			& 2\,811 						&& $<1$				& 708								&& $<1$		& 227\\
22 	& $<1$ 			& 11\,615 					&& $<1$				& 1\,142						&& $<1$		& 3\,237\\
24 	& $<1$ 			& 12\,623 					&& $<1$				& 5\,332						&& $<1$		& 736\\
26 	& $<1$ 			& 37\,708 					&& $<1$				& 5\,313						&& $<1$		& 2\,311\\
28 	& $<1$ 			& 35\,530 					&& $<1$				& 16\,365						&& $<1$		& 85\,315\\
30 	& $<1$ 			& 650\,811 					&& $<1$				& 49\,620						&& $<1$		& 68\,033\\
32 	& $<1$ 			& 332\,306 					&& $<1$				& 91\,094						&& $<1$		& 22\,407\\
34 	& $<1$ 			& 1\,342\,216 			&& $<1$				& 131\,169					&& $<1$		& 21\,696\\
36 	& $<1$ 			& 2\,160\,102 			&& $<1$				& 524\,491					&& $<1$		& 248\,184\\
38 	& 5.34 			& 13\,469\,359 			&& $<1$				& 763\,317					&& $<1$		& 83\,636\\
40 	& 6.25 			& 16\,393\,039 			&& 1.70				& 7\,335\,775				&& $<1$		& 220\,480\\
42 	& 107.69		& 256\,686\,929			&& 2.74				& 11\,575\,637			&& $<1$		& 612\,423\\
44 	& 876.91 		& 1\,944\,525\,360 	&& 19.80			& 79\,587\,812			&& 1.02		& 2\,489\,017\\
46 	& 1\,573.31	& 3\,565\,703\,651	&& 10.22			& 38\,865\,293			&& 1.59 	& 3\,430\,033\\
48 	& 542.79 		& 1\,231\,902\,706	&& 1\,112.55	& 4\,289\,081\,568	&& 5.69		& 12\,080\,931\\
50 	& 6\,418.52 & --								&& 4\,018.20	& --								&& 17.38	& 34\,639\,665\\
\hline
\end{tabular}
\end{small}
\end{center}
\end{table}

\begin{sidewaystable}
\begin{center}
\caption{Solving all prob026 instances when $50 < T \le 70$. }\label{CTS3vsCTS2}
\begin{tabular}{crrcrrcrr}
\noalign{\smallskip}\hline
\multirow{2}{*}{$T$}
		& \multicolumn{2}{c}{\texttt{EnASS}$_2$ \cite{Hamiez&Hao2008}} &
		& \multicolumn{2}{c}{\texttt{EnASS}$_3$ (Sect.~\ref{CTS60})} & 
		& \multicolumn{2}{c}{\texttt{EnASS}$_4$ (Sect.~\ref{CTS70})}\\
			\cline{2-3} \cline{5-6} \cline{8-9}
	&Time 		&$|$BT$|$				 &&Time 			&$|$BT$|$ 				&&Time						&$|$BT$|$\\
\hline
52&-- 			&--							 &&377.84			&1\,345\,460\,512	&&50.11						&101\,432\,823\\
54&10.59 		&29\,767\,940		 &&763.08			&2\,802\,487\,580	&&101.74					&196\,808\,595\\
56&-- 			&--							 &&2\,552.65	&--								&&334.26					&753\,747\,164\\
58&269.88		&827\,655\,311	 &&13\,715.33	&--								&&878.96					&1\,851\,547\,682\\
60&-- 			&-- 						 &&198\,250.44&--								&&2\,364.47				&--\\
62&279.38		&494\,071\,117	 &&--					&--								&&9\,866.51				&--\\
64&--				&--							 &&--					&--								&&32\,386.67			&--\\
66&7\,508.51&1\,614\,038\,658&&--					&--								&&85\,989.73			&--\\
68&--				&								 &&--					&--								&&518\,194.31			&--\\
70&8\,985.05&--							 &&--					&--								&&1\,512\,574.41	&--\\
\hline
\end{tabular}
\end{center}
\end{sidewaystable}

From Table~\ref{CTS3vsCTS1}--\ref{CTS3vsCTS2}, one observes that \texttt{EnASS}$_3$ solves more prob026 instances than \texttt{EnASS}$_1$ within 3 hours. Indeed, while \texttt{EnASS}$_1$ is limited to $T \le 50$, \texttt{EnASS}$_3$ finds solutions for $T$ up to 56 in at most 67 minutes (see the $T = 50$ case in Table~\ref{CTS3vsCTS1}). Moreover, except two cases ($T \in \left\{ 16, 48\right\}$), the number of backtracks required to find a solution is much smaller for  \texttt{EnASS}$_3$  than for \texttt{EnASS}$_1$. 

Table~\ref{CTS3vsCTS2} shows that the comparison between \texttt{EnASS}$_3$ and \texttt{EnASS}$_2$ is somewhat mitigated. Indeed, \texttt{EnASS}$_3$ is able to find solutions for \textbf{all} $T$ up to 56 within 3 hours while \texttt{EnASS}$_2$ solves the instances up to $T=70$, but only when $T \bmod 4 \neq 0$. For the cases that are solved by both \texttt{EnASS}$_3$ and \texttt{EnASS}$_2$, \texttt{EnASS}$_2$ finds a solution much faster. On the other hand, \texttt{EnASS}$_3$ finds solutions for $T \in \left\{52, 56, 60 \right\}$ for which \texttt{EnASS}$_2$ fails. Finally, one notices that \texttt{EnASS}$_3$ requires much more time to solve the $T \in \left\{58, 60 \right\}$ instances (about 55 hours for $T = 60$).

\section{Solving all prob026 instances when $50 < T \le 70$}\label{CTS70}

The rule $r'_I$ used to solve \textbf{all} prob026 instances for $50 < T \le 70$ is similar to the original $r_I$ requirement introduced in \cite[Sect.~7]{Hamiez&Hao2008}. Indeed, like $r_I$, $r'_I$ inverses two weeks and keeps them invariant during the search. The only difference between $r_I$ and the new $r'_I$ rule is the weeks that are concerned: While $r_I$ considers weeks 2 and $W$, the $r'_I$ constraint inverses weeks 2 and $W-1$. More formally, $\forall w \in \{2, W-1\}, r'_I(w) \Leftrightarrow \forall\,1 \le p \le P, \langle p,w \rangle = \overline{s}\langle P-p+1,w \rangle$.

This leads to \texttt{EnASS}$_4$ which comes from \texttt{EnASS}$_3$ by adding in $\mathcal{R}_3$ the new $r'_I$ rule: $\mathcal{R}_4 = \{c_\mathcal{P}, c_\mathcal{D}, r'_\Rightarrow, r'_I\}$. Since the first two weeks are now invariant (and the last two due to $r'_{\Rightarrow}$), $w_f$ has to be set to 3 before running \texttt{EnASS}$_4$. Table~\ref{Example_valid_schedule} in Sect.~\ref{SectIntroduction} shows an example of a solution found by \texttt{EnASS}$_4$ (and \texttt{EnASS}$_3$) for $T=8$: For instance, the first match in week 2 is $\overline{s}\langle 4-1+1,2 \rangle$, i.e., $\langle 1, 2 \rangle = (6, 8)$.

The computational performance of the \texttt{EnASS}$_4$ variant is provided in Table~\ref{CTS3vsCTS1} for $6 \le T \le 50$ and in Table~\ref{CTS3vsCTS2} for $50 < T \le 70$\footnote{The first solution found by \texttt{EnASS}$_4$ for $50 < T \le 70$ is available on-line from \texttt{http://www .info.univ-angers.fr/pub/hamiez/EnASS4/Sol52-70.html}.}. One notices that \texttt{EnASS}$_4$ is faster than \texttt{EnASS}$_3$ and \texttt{EnASS}$_1$ (see the ``$|$BT$|$'' columns in Table~\ref{CTS3vsCTS1}) to solve instances when $T \ge 12$ (and for $T=8$), except for the $T \in \{12, 14, 16, 22, 28, 30\}$ cases. Furthermore, within 3 hours per $T$ value, \texttt{EnASS}$_4$ is capable of solving larger instances (up to $T=62$, see Table~\ref{CTS3vsCTS2}) than \texttt{EnASS}$_1$ ($T \le 50$) and \texttt{EnASS}$_3$ ($T \le 56$). While \texttt{EnASS}$_2$ solves only some instances for $50 < T \le 70$ (those verifying $T \bmod 4 \neq 0$, see Table~\ref{CTS3vsCTS2}), \texttt{EnASS}$_4$ finds solutions for all these cases. This is achieved within 3 hours for $T$ up to 62, but larger instances can require more execution time (about 18 days for $T=70$). Finally, note that adding the new $r'_I$ rule excludes solutions for $T \in \{6, 10\}$. 

\section{Conclusion}

We provided in this short note two enhancements to an \texttt{En}umerative \texttt{A}lgo\-rithm for \texttt{S}ports \texttt{S}cheduling (\texttt{EnASS}) previously proposed in \cite{Hamiez&Hao2008}. These enhancements are based on additional properties (identified in \emph{some} solutions) as new constraints to reduce the search tree constructed by the algorithm. With these enhancements, all prob026 instances with $T \leq 70$ can be solved for the first time. Since the main idea behind the enhancements is to add refined requirement rules in the \texttt{EnASS} method, we expect that the method can be further improved to solve prob026 instances for $T > 70$.

\section*{Acknowledgments}
This work was partially supported by the ``Pays de la Loire'' Region (France) within the LigeRO (2010 -- 2013) and RaDaPop (2009 -- 2013) projects.

\end{document}